\documentclass{Rinton-P9x6}

\begin{document}

\title{Forms on Vector Bundles Over Compact Hyperbolic Manifolds
and Entropy Bounds}

\author{A. A. Bytsenko}

\address{Departamento de F\'{\i}sica, Universidade Estadual de Londrina, 
Londrina, Brazil
}

\author{V. S. Mendes}

\address{Departamento de F\'{\i}sica, Universidade Estadual de Londrina,  
Londrina, Brazil
}

\author{A. C. Tort}

\address{Departamento de F\'{\i}sica Te\'orica, Instituto de F\'{\i}sica Universidade
Federal do Rio de Janeiro,
Rio de Janeiro, Brazil
}



\maketitle

\abstracts{
We analyze gauge theories based on abelian $p-$forms in real compact
hyperbolic manifolds. The explicit thermodynamic functions
associated with skew--symmetric tensor fields are obtained via zeta--function regularization and the trace tensor kernel
formula. Thermodynamic quantities in the high--temperature expansions 
are calculated and the entropy/energy ratios are established.}

\section{Introduction}

It is known that the thermodynamics of quantum
fields in an Einstein universe for some radius is equivalent to that
of an instantaneously static closed Friedmann--Robertson--Walker
universe. The field
thermodynamics of positive curvature Einstein spaces was discussed
by several authors before. In particular, the so-called entropy
bounds or entropy to thermal energy ratios were calculated and
compared with known bounds such as the Bekenstein bound or the
Cardy-Verlinde bound. For example, for a massless scalar field in
${\bf S}^3$ space this was done in \cite{Breviketal2003} and for a
massive scalar field in \cite {Elizalde&Tort2003}. Here we
wish to extend the evaluation of those type of bounds to the case
of skew symmetric tensor fields in real hyperbolic spaces.

\section{Quantum Dynamics of Exterior Forms of Real Hyperbolic Spaces}

We shall work with a $D-$dimensional compact hyperbolic space $X$
with universal covering $M$ and fundamental group $\Gamma$. We can represent
$M$ as the symmetric space $G/K$, where $G=SO_1(D,1)$ and $K=SO(D)$ is a
maximal compact subgroup of $G$. Then we regard $\Gamma$ as a discrete
subgroup of $G$ acting isometrically on $M$, and we take $X$ to be the
quotient space by that action: $X=\Gamma\backslash M= \Gamma\backslash G/K$.
Let $\tau$ be an irreducible representation of $K$ on a complex vector space
$V_\tau$, and form the induced homogeneous vector bundle $G\times_K V_\tau$
(the fiber product of $G$ with $V_\tau$ over $K$) $\rightarrow M$ over $%
M $. Restricting the $G$ action to $\Gamma$ we obtain the quotient bundle $%
E_\tau=\Gamma\backslash (G\times_KV_\tau)\rightarrow X=\Gamma\backslash
M $ over $X$. The natural Riemannian structure on $M$ (therefore on $X$)
induced by the Killing form $(\;,\;)$ of $G$ gives rise to a connection
Laplacian ${\cal L}$ on $E_\tau$. If $\Omega_K$ denotes the Casimir
operator of $K$ -- that is
$\Omega_K=-\sum y_j^2$,
for a basis $\{y_j\}$ of the Lie algebra ${k}_0$ of $K$, where 
$(y_j\;,y_\ell)=-\delta_{j\ell}$, then $\tau(\Omega_K)=\lambda_\tau{}$ for a
suitable scalar $\lambda_\tau$. Moreover for the Casimir operator $\Omega$
of $G$, with $\Omega$ operating on smooth sections $\Gamma^\infty E_\tau$ of
$E_\tau$ one has
$
{\cal L}=\Omega-\lambda_\tau {\bf 1}\;;
$
see Lemma 3.1 of \cite{Wallach}. For $\lambda\geq 0$ let
\begin{equation}  \label{03}
\Gamma^\infty\left(X\;,E_\tau\right)_\lambda= \left\{s\in\Gamma^\infty
E_\tau\left|-{\cal L}s=\lambda s\right. \right\}
\end{equation}
be the space of eigensections of ${\cal L}$ corresponding to $\lambda$.
Here we note that since $X$ is compact we can order the spectrum of $-{\cal 
L}$ by taking $0=\lambda_0<\lambda_1<\lambda_2<\cdots$; $\lim_{j\rightarrow
\infty}\lambda_j=\infty$. It will be
convenient moreover to work with the normalized Laplacian ${\cal L}_p=-c(D)
{\cal L}$ where $c(D)=2(D-1)=2(2N-1)$. ${\cal L}_p$ has spectrum 
$\left\{c(D)\lambda_j\;,m_j\right\}_{j=0}^\infty$ where the multiplicity $m_j$
of the eigenvalue $c(D)\lambda_j$ is given by
$
m_j={\rm dim}\;\Gamma^\infty\left(X\;,E_{\tau^{(p)}}\right)_{\lambda_j}
$.

It is easy to prove the following properties for
operators and forms: $dd=\delta\delta=0$,\, $\delta = (-1)^{Dp+D+1}*d*$,\, 
$**\omega_p = (-1)^{p(D-p)}\omega_p$. Let $\alpha_p,\, \beta_p$ be $p-$forms;
then the invariant inner product is defined by $(\alpha_p, \beta_p):=
\int_M \alpha_p\wedge*\beta_p$. The operators $d$ and $\delta$ are
adjoint to each other with respect to this inner product for $p-$forms: $%
(\delta\alpha_p, \beta_p) = (\alpha_p, d\beta_p)$.
In quantum field theory the Lagrangian associated with $\omega_p$ takes the
form: $L=d\omega_p\wedge *d\omega_p$ (gauge field);\, $L=\delta\omega_p%
\wedge*\delta\omega_p$ (co--gauge field). The Euler--Lagrange equations
supplied with the gauge give ${\cal L}_p\omega_p =0\,,\,\,\delta\omega_p =0$
(Lorentz gauge);\, ${\cal L}_p\omega_p =0\,,\,\, d\omega_p =0$ (co--Lorentz
gauge). These Lagrangians give possible representation of tensor fields or
generalized Abelian gauge fields. The two representations of tensor fields
are not completely independent. Indeed, there is a duality property in the
exterior calculus which gives a connection between star--conjugated gauge
tensor fields and co--gauge fields.

\section{The Trace Formula Applied to the Tensor Kernel}

We can apply the version of the trace
formula developed by Fried in \cite{Fried}. First we define additional
notation. For $\sigma_p$ the natural representation of $SO(2N-1)$ on $%
\Lambda^p {\bf C}^{2N-1}$ one has the corresponding
Harish--Chandra--Plancherel density given, for a suitable normalization of
Haar measure $dx$ on $G$, by
\begin{equation}  \label{07}
\mu_{\sigma_p(r)}= \frac{\pi}{2^{4k-4}[\Gamma(N)]^2} \left(
\begin{array}{c}
2N-1 \\
p
\end{array}
\right) P_{\sigma_p}(r) r\tanh(\pi r)\;,
\end{equation}
for $0\le p \le N-1$, where
\begin{equation}  \label{08}
P_{\sigma_p}(r) =  \prod_{\ell=2}^{p+1} \left[ r^2+\left(N-\ell+\frac{3}{2}
\right)^2 \right] 
\prod_{\ell=p+2}^{N} \left[ r^2+\left(N-\ell+\frac{1}{2}
\right)^2 \right]\;
\end{equation}
is an even polynomial of degree $2N-2$. One has that $P_{\sigma_p}(r)=
P_{\sigma_{2N-1-p}}(r)$ and $\mu_{\sigma_p}(r)=\mu_{\sigma_{2N-1-p}}(r)$ for
$N\le p\le 2N-1$. Now define the Miatello coefficients \cite{Miatello} 
$a_{2\ell}^{(p)}$ for $G=SO_1(2N+1, 1)$ by
$
P_{\sigma _{p}}(r)=\sum_{\ell =0}^{N-1}a_{2\ell }^{(p)}r^{2\ell }\;,\qquad
0\leq p\leq 2N-1\;.  \label{09}
$
Let ${\rm Vol}(\Gamma \backslash G)$ denote the integral of the
constant function ${\bf 1}$ on $\Gamma \backslash G$ with respect to the $G-$
invariant measure on $\Gamma \backslash G$ induced by $dx$.
For $0\leq p\leq D-1$ the Fried trace formula \cite{Fried}
applied to the tensor kernel associated with co--exact forms 
has to be modified \cite{Bytsenko97,Bytsenko03}:
\begin{eqnarray}
{\rm Tr}\left( e^{-t{\cal L}_{p}}\right) &=&\sum_{j=1}^{p}\left(
-1\right) ^{j}\left[ I_{\Gamma }^{(p-j)}({\cal K}_{t})+I_{\Gamma }^{(p-1-j)}
({\cal K}_{t})\right.  \nonumber \\
&&+\left. H_{\Gamma }^{(p-j)}({\cal K}_{t})+H_{\Gamma }^{(p-1-j)}({\cal K}
_{t})-b_{p-j}\right]
\mbox{,}
\end{eqnarray}
where $b_{p}$ are the Betti numbers.
In the above formula 
$I_{\Gamma }^{(p)}({\cal K}_{t}),H_{\Gamma }^{(p)}({\cal K}_{t})$ are the
identity and hyperbolic orbital integral respectively.

\section{The Spectral Functions of Exterior Forms}

The spectral zeta function related to the Laplace operator ${\cal L}_{j}$
can be represented by
the inverse Mellin transform of the heat kernel 
${\mathcal K}_t = {\rm Tr}\,\exp \left( -t {\cal L}_{j}\right)$.
Using the Fried formula, we can write the
zeta function as a sum of two contributions:
\begin{eqnarray}
\zeta ( s|{\cal L}_{j}) 
& = &\frac{1}{\Gamma (s)}\int_{0}^{\infty }dtt^{s-1}\left( I_{\Gamma
}^{(j)}(\mathcal{K}_{t})+I_{\Gamma}^{(j-1)}(\mathcal{K}_{t})
+ H_{\Gamma }^{(j)}(\mathcal{K}_{t})+H_{\Gamma }^{(j-1)}(\mathcal{K}
_{t})\right) \nonumber \\
& \equiv & \zeta _{I}^{(D)}(s,j)+\zeta _{H}^{(D)}(s,j)
\mbox{.}
\end{eqnarray}
For the identity component we have
\begin{equation}
\zeta _{I}^{(D)}(s,j)=\frac{V_{\Gamma}}{\Gamma (s)}\int_{0}^{\infty }dt
t^{s-1} \int_{\bf{R}}dr\,\mu _{\sigma
_{j}}e^{-t(r^{2}+\alpha _{j}^{2})},
\end{equation}
where $V_{\Gamma }=\chi (1){\rm Vol}\left( \Gamma \backslash G\right)/4\pi$,
and we define
$\alpha _{j}^{2}=b^{(j)}+ (\rho_{0}-j)^{2}$, $\rho_0=(D-1)/2$
and $b^{(j)}$ are constants.
Replacing the Harish--Chandra--Plancherel measure, we obtain two
representations for $\zeta ^{(D)}_{I}(s,j)$ which hold for 
odd and even dimension:
\begin{eqnarray}
&& \zeta_{I}^{(2N)}(s,j) = \frac{V_{\Gamma }C_{2N}^{(j)}}{\Gamma
(s)}\sum_{\ell =0}^{N-1}a_{2\ell ,2N}^{(j)}\int_{0}^{\infty}dt\,t^{s-1}
\int_{\bf{R}}drr^{2\ell +1}\mathrm{tanh}(\pi
r)\,e^{-t(r^{2}+\alpha_{j}^{2})} \nonumber \\
&& = \frac{V_{\Gamma }C_{2N}^{(j)}}{\Gamma
(s)}\sum_{\ell =0}^{N-1}a_{2\ell ,2N}^{(j)} 
\left[ \frac{\Gamma (\ell +1)\Gamma (s-\ell -1)}
{\alpha_{j}^{2s-2\ell -2}} 
+ \sum_{n=0}^{\infty }\xi _{n\ell }
\frac{\Gamma (s+n)}{\alpha_{j}^{2s+2n}}\right],
\label{zeta2}
\end{eqnarray}
\begin{eqnarray}
&& \zeta_{I}^{(2N+1)}(s,j) = \frac{V_{\Gamma }C_{2N+1}^{(j)}}
{\Gamma (s)}\sum_{\ell =0}^{N}a_{2\ell ,2N+1}^{(j)} 
\int_{0}^{\infty }dt\,t^{s-1}\int_{\bf{R}}drr^{2\ell}
e^{-t(r^{2}+\alpha _{j}^{2})} \nonumber \\
&& = \frac{V_{\Gamma }C_{2N+1}^{(j)}}{
\Gamma (s)}\sum_{\ell =0}^{N}\Gamma \left( \ell
+\frac{1}{2}\right) 
\Gamma \left( s-\ell -\frac{1}{2}\right) a_{2\ell, 2N+1}^{(j)}
\alpha _{j}^{-2s+2\ell +1}
\mbox{,}
\end{eqnarray}
where $B_{n}$ is the $n-$th Bernoulli number, and where we define
\begin{equation}
\xi _{n\ell } : = \frac{\left( -1\right) ^{\ell +1}\left(
1-2^{-2\ell -2n-1}\right) }{n!\left( 2\ell +2n+2\right) }B_{2\ell +2n+2}.
\label{csi}
\end{equation}
In fact we do not need the hyperbolic component $\zeta_{H}^{(D)}(s,j)$
since at high temperature expansion (see next section) only the function $\zeta_{I}^{(D)}(s,j)$ will be used.

\section{The High Temperature Expansion}

Using the Mellin representation for the zeta function one can
obtain useful formulae for the non--trivial temperature dependent
part of the identity and hyperbolic orbital components of the
free energy (for details see \cite{Bytsenko1,Bytsenko2,Bytsenko3})
\begin{equation}
F_{I,H}^{(D)}(\beta,j)=-\frac{1}{2\pi i}\int\limits_{\Re
z=c}dz\,\zeta _{R}(z)\Gamma (z-1)\zeta _{I,H}^{(D)}
((z-1)/2, j) \beta^{-z}
\mbox{.}
\end{equation}
A tedious calculation gives the following result:
\begin{eqnarray}
F_{I}^{(2N)}(\beta,j) & = &-\frac{V_{\Gamma }C_{2N}^{(j)}a_{2N-2,2N}^{(j)}
}{\sqrt{4\pi }}\Gamma (N)\zeta (2N+1)  
\Gamma \left( N+\frac{1}{2}\right) \beta ^{-2N-1}  \nonumber \\
&&-\frac{V_{\Gamma }C_{2N}^{(j)}}{\sqrt{4\pi }}\zeta (2N-1)\Gamma \left( N-
\frac{1}{2}\right)  
\left[ a_{2N-4,2N}^{(j)}\Gamma (N-1)\right.  \nonumber \\
&&-\left. a_{2N-2,2N}^{(j)}\Gamma \left( N\right) \right] \beta^{-2N+1}
+{\cal O}(\beta ^{-2N+3}),
\end{eqnarray}
\begin{eqnarray}
F_{I}^{(2N+1)}(\beta,j) & = &-\frac{V_{\Gamma
}C_{2N+1}^{(j)}a_{2N,2N+1}^{(j)}}{\sqrt{4\pi }}\Gamma \left( N+\frac{1}{2}
\right)
\zeta (2N+2)\Gamma (N+1)\beta ^{-2N-2}  \nonumber \\
&&-\frac{V_{\Gamma }C_{2N+1}^{(j)}}{\sqrt{4\pi }}\zeta (2N)\Gamma
\left(
N\right) 
\left[ a_{2N-2,2N+1}^{(j)}\Gamma \left(
N-\frac{1}{2}\right)
\right.   \nonumber \\
&&-\left. a_{2N,2N+1}^{(j)}\Gamma \left( N+\frac{1}{2}\right)
\alpha_{j}^{2}\right] \beta ^{-2N}  
+{\cal O}\left( \beta ^{-2N+2}\right) .
\end{eqnarray}
The contribution associated to the hyperbolic orbital
component is negligible small. 

\subsection{The Thermodynamic Functions and the Entropy Bound}

In the context of the Hodge theory, the
physical degrees of freedom are represented by the co--exact forms. 
Thus the free energy becomes
\begin{equation}
{\cal F}^{(D)}(\beta )=\sum_{j=0}^{p}(-1)^{j}\left(
F^{(D)}_{I}(\beta,j)+F^{(D)}_{I}(\beta,j)\right) 
\mbox{.}
\end{equation}
In the high temperature limit $(\beta \rightarrow 0$) we have
\begin{equation}
{\cal F}^{(D)}(\beta ) = -A_{1}(D;\Gamma )\beta^{-D-1}
-A_{2}(D;\Gamma )\beta^{-D+1}+{\cal O}(\beta^{-D+3})
\mbox{,}
\end{equation}
where for the even dimensional case
\begin{eqnarray}
A_{1}\left( 2N;\Gamma \right) & = &\frac{V_{\Gamma }}{\sqrt{4\pi
}}\zeta (2N+1) \Gamma \left( N\right)\Gamma \left(N+\frac{1}{2}\right) 
\sum_{j=0}^{p}(-1)^{j}\left( C_{2N}^{\left(
p-j\right)}a_{2N-2,2N}^{\left( p-j\right) }\right.  \nonumber \\
& + &
\left. C_{2N}^{(p-1-j)}a_{2N-2,2N}^{(p-1-j)}\right), 
\label{a}
\end{eqnarray}
\begin{eqnarray}
A_{2}(2N;\Gamma ) & = &\frac{V_{\Gamma }}{\sqrt{4\pi }}\zeta
(2N-1)\Gamma \left( N-\frac{1}{2}\right) \sum_{j=0}^{p}(-1)^{j}
\left[\Gamma (N-1)\left(
C_{2N}^{(p-j)}a_{2N-4,2N}^{(p-j)}\right. \right.  \nonumber \\
& + &
\left. C_{2N}^{(p-1-j)}a_{2N-4,2N}^{(p-1-j)}\right)  
+\Gamma (N)\left(C_{2N}^{(p-j)}a_{2N-2,2N}^{(p-j)}\alpha_{p-j}^{2}
\right. \nonumber \\  
& + &
\left. \left. C_{2N}^{(p-1-j)}a_{2N-2,2N}^{(p-1-j)}
\alpha_{p-1-j}^{2}\right) \right]
\mbox{,}  
\label{b}
\end{eqnarray}
and for the odd dimensional case
\begin{eqnarray}
A_{1}(2N+1;\Gamma )  & = & \frac{V_{\Gamma }}{\sqrt{4\pi }}
\zeta (2N+2)\Gamma \left( N+\frac{1}{2}\right) \Gamma (N+1)
\sum_{j=0}^{p}(-1)^{j}\left(
C_{2N+1}^{(p-j)}a_{2N,2N+1}^{(p-j)}\right.  \nonumber \\
& + & \left. C_{2N+1}^{(p-1-j)}a_{2N,2N+1}^{(p-1-j)}\right) \label{c}
\end{eqnarray}
\begin{eqnarray}
A_{2}(2N+1;\Gamma ) & = &\frac{V_{\Gamma }}{\sqrt{4\pi }}\zeta
(2N)\Gamma
(N)\sum_{j=0}^{p}(-1)^{j} 
\left[\Gamma \left( N-\frac{1}{2}\right) \left(
C_{2N+1}^{(p-j)}a_{2N-2,2N+1}^{(p-j)}\right. \right.   \nonumber \\
& + & \left. C_{2N+1}^{(p-1-j)}a_{2N-2,2N+1}^{(p-1-j)}\right)  
-\Gamma \left( N+\frac{1}{2}\right) \left(
C_{2N+1}^{(p-j)}a_{2N,2N+1}^{(p-j)}\alpha _{p-j}^{2}\right.   \nonumber \\
& + &
\left. \left. C_{2N+1}^{(p-j-1)}a_{2N,2N+1}^{(p-j-1)}\alpha
_{p-1-j}^{2}\right) \right]
\mbox{.}  
\label{d}
\end{eqnarray}
In fact, in the sums (\ref{a}) - (\ref{d}) only terms containing the Miatello
coefficients $a_{2\ell, D}^{(p)}$ survive and define the coefficients
$A_1$ and $A_2$.
The entropy and the total energy can be obtained with the help of the
following thermodynamic relations: $S^{(D)}(\beta )=\beta ^{2}{\partial
}{\cal F}^{(D)}(\beta )/\partial \beta $, $E^{(D)}(\beta )={\partial
}(\beta {\cal F}^{(D)}(\beta ))/{\partial }\beta $. Therefore,
\begin{equation}
S^{(D)}(\beta ) = (D+1)A_{1}(D;\Gamma)
\beta ^{-D}+ (D-1)A_{2}(D;\Gamma) \beta^{-D+2}+ 
{\cal O}\left( \beta ^{-D+4}\right) 
\mbox{,}
\end{equation}
\begin{equation}
E^{(D)}(\beta ) = -DA_{1}(D;\Gamma ) \beta ^{-D-1}-(D-2)A_{2}(D;\Gamma) 
\beta^{-D+1}+{\cal O}\left( \beta ^{-D+3}\right)
\mbox{.}
\end{equation}
The entropy/energy ratio becomes
\begin{equation}
\frac{S^{(D)}(\beta )}{E^{(D)}(\beta )}=\frac{D+1}{D}\beta 
+\frac{2}{D^{2}}\frac{
A_{2}(D;\Gamma )}{A_{1}(D;\Gamma )}\beta ^{3}+{\cal O}\left( \beta
^{5}\right) .
\label{bound}
\end{equation}

\section{Conclusions}

We have obtained the high--temperature expansion for the entropy/energy ratios 
of abelian gauge fields in real compact hyperbolic spaces. 
The dependence on the Miatello coefficients related to the structure of the
Harish--Chandra--Plancherel measure starts from the second term of the
expansion. In the case of scalar fields $(p=0)$ we have Eq. (\ref{bound}) 
with
\begin{equation}
\frac{A_{2}(2N;\Gamma )}{A_{1}(2N;\Gamma )}=\frac{2}{2N-1}\frac{\zeta (2N-1)
}{\zeta (2N+1)}\left( \frac{1}{N-1}\frac{a_{2N-4,2N}^{(0)}}{a_{2N-2,2N}^{(0)}
}-\alpha _{0}^{2}\right) ,
\end{equation}
\begin{equation}
\frac{A_{2}\left( 2N+1;\Gamma \right) }{A_{1}\left( 2N+1;\Gamma \right) }=
\frac{1}{N}\frac{\zeta \left( 2N\right) }{\zeta \left( 2N+2\right)
}\left( \frac{2}{2N-1}\frac{a_{2N-2,2N+1}^{\left( 0\right)
}}{a_{2N,2N+1}^{\left( 0\right) }}-\alpha _{0}^{2}\right) ,
\end{equation}
where $\alpha _{0}^{2}=\rho _{0}^{2}+m^{2}$ 
($\alpha _{0}^{2}=\rho _{0}^{2}$ for the massless case).
For three--dimensional hyperbolic manifolds the Miatello coefficients 
reads \cite{Bytsenko003}:
$a_{0}^{(0)}=a_{2}^{(0)}=1$ and therefore
$
S^{(3)}(\beta )/E^{(3)}(\beta ) = (4/3)\beta + (10/3\pi ^{2})
(2-\alpha _{0}^{2})\beta ^{3}+{\cal O}(\beta ^{5})
$.
This formula is in agreement with result obtained in 
\cite{Elizalde&Tort2003}
where entropy bounds have been calculated for spherical geometry
and where the dependence on the geometry of the background also starts from 
the second term of the expansion.


\end{document}